\documentclass[iop]{emulateapj}

\slugcomment{2018.9.19}

\shorttitle{Phase spirals}
\shortauthors{Wang C. et al.}

\begin{document}


\title{The Galactic  disc phase spirals at different Galactic positions revealed by Gaia and LAMOST data}


\author{C. Wang\altaffilmark{1,8,9},  Y. Huang\altaffilmark{2},  H.-B. Yuan\altaffilmark{3},  M.-S. Xiang\altaffilmark{4,5}, B.-Q. Chen\altaffilmark{2}, H.-F. Wang\altaffilmark{2,6,9}, Y.-Q. Wu\altaffilmark{5},  H.-W. Zhang\altaffilmark{1},  Z.-J. Tian\altaffilmark{7}, Y. Yang\altaffilmark{2}, M. Zhang\altaffilmark{1}, X.-W. Liu\altaffilmark{2,8}
}

\altaffiltext{1}{Department of Astronomy, Peking University, Beijing 100871, People's Republic of China.}
\altaffiltext{2}{South-Western Institute for Astronomy Research, Yunnan University, Kunming, Yunnan 650091, People's Republic of China.}
\altaffiltext{3}{Department of Astronomy, Beijing Normal University, Beijing 100875, People's Republic of China.}
\altaffiltext{4}{Max-Planck Institute for Astronomy, Konigstuhl, D-69117, Heidelberg, Germany.}
\altaffiltext{5}{National Astronomical Observatories, Chinese Academy of Sciences, Beijing 100012, People's Republic of China.}
\altaffiltext{6}{Department of Astronomy, China West Normal University, Nanchong 637009, People's Republic of China.}
\altaffiltext{7}{Department of Astronomy, Yunnan University, Kunming 650200, People's Republic of China.}
\altaffiltext{8}{E-mails: wchun@pku.edu.cn (CW); x.liu@ynu.edu.cn (XWL)}
\altaffiltext{\altaffilmark{9}}{LAMOST FELLOW}

\begin{abstract}
We have investigated the distributions of stellar azimuthal and radial velocity components $V_{\Phi}$ and $V_{R}$ in the vertical position-velocity plane  $Z$-$V_{Z}$ across the Galactic disc of $6.34 \lesssim R \lesssim 12.34$\,kpc and $|\Phi| \lesssim 7.5^{\circ}$ using a Gaia and  Gaia-LAMOST sample of stars.  
As found in previous works, the distributions exhibit significant spiral patterns. The  $V_{R}$ distributions also show clear quadrupole patterns, which are the consequence of the well-known tilt of the velocity ellipsoid.  
The observed  spiral and quadrupole patterns in the phase space plane vary strongly with radial and azimuthal positions. 
The phase spirals of $V_{\Phi}$  become more and more relaxed as $R$ increases. 
The spiral patterns of $V_{\Phi}$ and $V_{R}$ and the quadrupole patterns of $V_{R}$  are strongest at $-2^{\circ} < \Phi < 2^{\circ}$ but negligible at $4^{\circ} < \Phi < 6^{\circ}$ and $-6^{\circ} < \Phi < -4^{\circ}$.  Our results suggest an external origin of the phase spirals. In this scenario, the intruder, most likely the previously well-known Sagittarius dwarf galaxy,  passed through the Galactic plane in the direction towards either Galactic center or anti-center.   The azimuthal variations of the phase spirals also help us constrain the   passage duration of the intruder.  A detailed model is required to reproduce the observed radial and azimuthal  variations of the phase spirals of $V_{\Phi}$ and $V_{R}$.

\end{abstract}


\section{Introduction}
It is recognized that  the Milky Way is not steady and axisymmetric.  
The stellar populations of the Milky Way disc are perturbed  by  non-axisymmetric structures, including the Bar, the Spiral arms,  the halo substructures and the satellite dwarf galaxies  \citep{siebert2012,gomez,bovy}, and therefore show significant phase spirals  \citep{antoja,joss,tian2018}, as well as radial motions and vertical bulk motions \citep{sun,williams,carlin,siebert2011,huang,carrillo,wang2018}.  
Studying those phase spirals and bulk motions  can help us understand the perturbation history of the Milky Way (disc). 


\cite{antoja} first detect the remarkable  phase spirals   in the local stellar disc.  
They find that the distributions of stellar radial and azimuthal velocity components $V_{R}$ and $V_{\Phi}$ show significant spiral patterns in  the vertical position-velocity plane $Z$-$V_{Z}$.  Further works have studied the variations of the phase spirals with  stellar age, action, chemistry and disc position  using theLAMOST-Gaia  \citep{tian2018} and  GALAH-Gaia  \citep{joss} data. Their results not only confirm the original remarkable discovery of  \cite{antoja}, but also resolve significant  variations of the phase spirals with aforementioned parameters.  \cite{tian2018} find that  the spirals are gradually apparent from $\tau <$\,0.5\,Gyr, and then slowly disappearing until $\tau\,>$\,6.0\,Gyr.
\cite{joss} show that the phase spirals are easier discerned  in the distributions of $\alpha$-poor stars than those of $\alpha$-rich stars.  They also find that the spiral is clearest in stars with smaller action $J_{\rm r}$, and tighter in stars with smaller angular momentum $L_{\rm z}$. The radial and azimuthal variations of phase spirals are also found by \cite{joss}.  It's worth noting that the sample of \cite{joss} only covers a small distance range of 1\,kpc from the Sun, the spatial variations of the phase spirals need to be explored  in a larger volume of the Galactic disc.

In the theoretical view, several works have attempted  to explain the observed phase space spiral patterns by either  external  \citep{antoja, joss, binney2018} or  internal  \citep{Khoperskov}  perturbations.  Both the two scenarios can reproduce the phase spirals of $V_{\Phi}$ and $V_{R}$  in the Solar neighborhood.   \cite{antoja}, \cite{binney2018}, and \cite{ joss}   suggest that  the phase spirals are probably the consequence of the Sagittarius  dwarf galaxy perturbation.   Independent to the external perturbation scenario,  \cite{Khoperskov} suggest that the observed phase space spirals can be produced naturally by vertical oscillations driven by the buckling of the stellar bar--no need of an  external perturber (a massive satellite or a sub-halo). Whereas, simulations considering external or internal perturbations predict different spatial variations of the phase spirals.  In the external scenario,  \cite{joss} imply tighter phase space spirals in the inner disc due to faster vertical oscillations leaded by a stronger disc gravity in the inner disc.  \cite{Khoperskov} have also predicted the properties of phase space spirals in different $R$ and $\Phi$ bins considering internal perturbations, but doesn't find a tighter phase space spirals in the inner disc.  
Thus, studying the  phase space spirals at different disc positions  in a larger volume of the disc can no doubt  distinguish the origins of the phase spirals.  Unfortunately,  hitherto all the  observations focus on the Solar neighbourhood. 


In the current work, we  study the phase space spirals at different disc  positions  ($R$ and $\Phi$), in particular its radial variations to see whether the phase space spirals in the inner disc are tighter as predicted by the scenario of  \cite{joss}.   Our work is based on the recently released Gaia DR2, which provides precise proper motions and distances for more than 1.3 billion stars, and precise line-of-sight velocities for more than 7 million stars.  In addition, the LAMOST surveys have yielded precise line-of-sight velocities and metallicities for millions of stars.  These data enable us to derive   accurate three dimensional velocities  for large samples of stars  across the Galactic disc ranging from 6\,kpc to 12\,kpc, thus allowing us  to examine the spatial variations of the phase space spirals across a wide range of the disc. 

This paper is  organised as follows.   In Section 2,  we briefly introduce the  samples used.  In Section 3, we  present the main results.  The discussions are presented in Section 4.   Finally, we summary our work in Section\, 5.


\section{Data}

\subsection{Coordinate systems}

We use the Galactocentric cylindrical system ($R, \Phi, Z$) with $R$, the projected
 Galactocentric  distance, increasing radially outwards, $\Phi$ in the direction of the Galactic rotation and $Z$ towards the North Galactic Pole.  
 The Sun is assumed to be at the Galactic mid-plane (i.e.\,$Z_{0}$\,$=$\,0\,pc)  and has a value of $ R_{0}$ of 8.34\,kpc  \citep{reid}.  We adopt a local circular speed of rotation 
curve of $V_{c}(R_{0})$\,$=$\,$240\, \rm \, km \, s^{-1}$, and solar motions ($U_{\odot},V_{\odot},W_{\odot}$)\,$=$\,($11.1,12.24,7.25$)\,$\rm km\,s^{-1}$ 
relative to the Local Standard of Rest  \citep{schonrich}.  The results presented in the current work are stable if we choose $Z_{0}$\,$=$\,27\,pc and other solar motions relative to the Local Standard of Rest \citep{Huang2015}. 

\subsection{The stellar samples}
By June 2016, $\sim$ 6.5 million stellar spectra of signal-to-noise ratios (SNRs) higher than 10 for 4.4 million unique stars  have been obtained with LAMOST  \citep{Xiang_msto} during the Pilot Surveys and the first four years of the five-year Phase I Regular Surveys of the LAMOST Galactic spectroscopic surveys  \citep{lamost, deng-legue, liu-lss-gac, yuan-lamost}. 
Stellar atmospheric  parameters   (effective temperature $T_{\mathrm{eff}}$, \, surface gravity $\log\,g$, metallicity $ \mathrm{ \, [Fe/H]} $)  and line-of-sight velocities $V_{l}$ with random error of 5\,$\rm km\,s^{-1}$ derived  from the spectra using LSP3  \citep{lsp3,kpca} are available. 

Precise parallaxes and proper motions of 1.3 billion stars in the Milky Way are now provided by the Gaia DR2  \citep{gaiadr2}.  Distances and asymmetric uncertainties are also available for those 1.3 billion stars in Gaia DR2  provided by \cite{bailer2018}, who  derive the distances and  uncertainties  with a Bayesian method. In the current work, we adopt their distances. Line-of-sight velocities $V_{l}$ for 7 million stars are also provided with an accuracy of $\sim$\,1\,$\rm km\,s^{-1}$. 
Combining the distances, proper motions and the line-of-sight velocities $V_{l}$ of millions of stars  provided by the Gaia DR2, we derive accurate three-dimensional velocities of 7,224,631 (hereafter named the "Gaia sample") stars. 
Three-dimensional velocities of 3,600,275 (hereafter named the "Gaia-LAMOST sample") stars are also derived using the distances, proper motions  from Gaia DR2 and $V_{l}$ from the LAMOST surveys.  


We estimate three-dimensional velocities ($U,V,W$) and associated uncertainties  of the individual stars  using the method of   \cite{johnson}. When we derive the errors of ($U,V,W$), the errors of distances, proper motions and line-of-sight velocities are considered based on the the principle  of  uncertainty propagation. Then we transform ($U,V,W$) to   ($V_{\rm R},V_{ \Phi},V_{\rm Z}$). Errors of  ($V_{\rm R},V_{ \Phi},V_{\rm Z}$) are also  estimated based on  the principle  of  uncertainty propagation.  

In order  to obtain  reliable results, we have removed stars 
with   $V_{\rm R},V_{ \Phi}$ and $V_{\rm Z}$ uncertainties larger than 50\,$\rm\,km^{-1}$, stars of distance uncertainties larger than 25\,per\,cent,  and stars of $|V_{\rm R}|\,>\,400\,\rm\,km\,s^{-1}$, or $|V_{\Phi}-240|\,>\,400\,\rm\,km\,s^{-1}$, or $|V_{\rm Z}|\,>\,400\,\rm\,km\,s^{-1}$. 
Finally, the Gaia sample contains 6,150,394 stars,  and the Gaia-LAMOST sample contains  3,344,860 stars.   

\section{Results} 
In this Section, we examine the spatial variations of the phase spirals of $V_{\Phi}$  and $V_{R}$  using the two samples. Firstly,  we divide the Gaia sample into 6  radial bins of $R$ from 6.34\,kpc to 11.34\,kpc and 6 azimuthal bins of $\Phi$ from $-6^{\circ}$ to $6^{\circ}$ to investigate the phase spirals of $V_{\Phi}$  and $V_{R}$ in the different bins of $R$ and $\Phi$.  Stars in the  Gaia-LAMOST sample are also  divided into 6 radial bins of $R$ from 7.34\,kpc to 12.34\,kpc to investigate the phase spirals of $V_{\Phi}$  and $V_{R}$ at different $R$.  The distribution of the Gaia-LAMOST sample stars in the $Z$-$\Phi$ plane is not symmetric, which is the consequence of the LAMOST sampling strategy, limiting magnitudes and so on.  Above the Galactic plane, most of the stars have negative values of $\Phi$, whereas below the Galactic plane, most of the stars have positive $\Phi$.   Thus, we do not study the azimuthal   variations of the phase spirals  with the Gaia-LAMOST sample.  
When we explore the radial and azimuthal variations of the phase spirals, we adopt an azimuthal range  $\Phi$ of [$-7.5^{\circ},7.5^{\circ}$]  and a radial range  $R$  of   [7.84, 8.84]\,kpc, respectively.    Likewise, when we  examine  the distributions of $V_{\Phi}$  and $V_{R}$ in the $Z$-$V_{Z}$ plane, we adopt bin  sizes of $\Delta Z = 0.01\,$kpc and $\Delta V_{Z} = 1.0\,\rm{km\,s^{-1}}$.

\subsection{The phase spirals at different Galactic positions as revealed by  the Gaia sample}
\subsubsection{Slicing by $R$}
\begin{figure}
\centering
\includegraphics[width=3.5in]{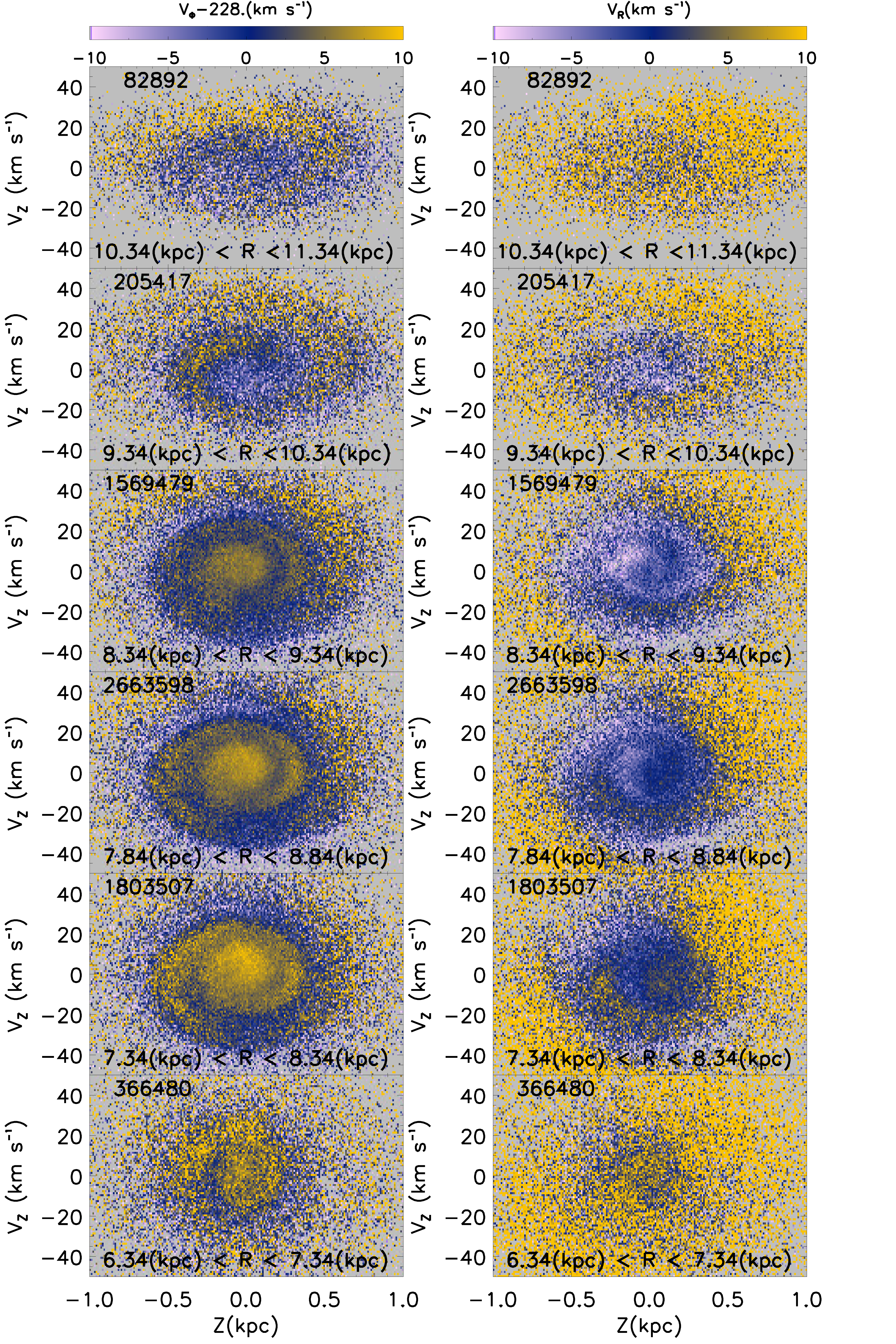}
\caption{ Distributions  of  median values of  $V_{\Phi}-228\,\rm{km s^{-1}}$ (left panels ) and $V_{R}$ (right panels) of the Gaia sample stars  in the $Z$-$V_{Z}$ plane  at different Galactic radii.  The radial range and number of stars of each bin are labeled in each of the panels.}
\label{low_alpha_r}
\end{figure}

We firstly investigate the radial variations of the phase space spirals   using the Gaia sample.  Fig.\,\ref{low_alpha_r} shows the distributions of $V_{\Phi}-228\,\rm{km s^{-1}}$ (left panels) and $V_{R}$ (right panels)  of the Gaia sample stars in the $Z$-$V_{Z}$ plane and in the different radial bins.  The radial range and number of stars in each bin are labeled in the Figure.  

The distributions of  $V_{\Phi}$ show significant phase spirals. Our results confirm the previous finding of   \cite{antoja},  \cite{joss} and  \cite{tian2018} for the Solar neighbourhood.  The phase spirals are apparent in all the radial bins, especially in the bin of $6.34\,\rm{kpc} <$ $R$ $< 10.34\,\rm{kpc}$. The  results strongly support the idea that the  phase spirals are a disc-wide phenomenon  \citep{joss}.   

Besides,  the phase spirals of $V_{\Phi}$  vary strongly with Galactic radius $R$ as Fig.\,\ref{low_alpha_r} shows.  The larger the $R$, the more relaxed of the phase spirals.  In Fig.\,\ref{low_alpha_r},  the inner spirals  become more and more relaxed as $R$ increases, and essentially  disappear at $R > $\,9.34\,kpc.  The outer spirals are stronger in the outer disc than in the inner disc. One can find 3 spirals at $R < $\,9.34\,kpc,  2 spirals at $9.34 < R < 10.34$\,kpc and only 1 spiral at $10.34 < R < 11.34$\,kpc  in the phase space plane  of $ -1 <  Z < 1$\,kpc and $-50 < V_{Z} < 50\, \rm{km\,s^{-1}}$.
The observed results presented here are consistent with the N-body simulation predictions of  \cite{joss} (see their Fig.\,22), who suggest that the stronger disc gravity in the inner disc leads to faster vertical oscillations and  tighter phase spirals. In this scenario, one can see more spirals in the inner disc than  in the outer disc in the same phase space  region, which is indeed what one sees in Fig.\,\ref{low_alpha_r}. 

The distributions of $V_{R}$  in the phase space plane  at different Galactic radii are also presented in Fig.\,\ref{low_alpha_r}. They show clear quadrupole patterns at all Galactic radii except that of $R > 10.34$\,kpc, values of $V_{R}$ are relatively small in the  top left and bottom right parts of each right panels compared to those in the top right and bottom left corners.  The quadrupole patterns, first mentioned by  \cite{joss} (see their Fig.\,16), are the consequence of the well-known tilt of the velocity ellipsoid  \citep{siebert2008, binney2014, joss}.  The quadrupole patterns are the clearest at 8.34\,kpc\,$< R <$\,9.34\,kpc with the smallest $V_{R}$  in the top left and the bottom right region, consistent with the previously observed radial velocity dip in the Solar neighborhood  \citep{siebert2011, huang, tianhaijun,carrillo}. 

At $7.34 < R <$\,10.34\,kpc, the phase spirals are also found in the distributions of $V_{R}$, which are broadly similar to those of $V_{\Phi}$.  However, the phase spirals are less tightly wound compared to those of $V_{\Phi}$, again consistent with the numerical simulation results  of  \cite{joss}. In the innermost ($R <$\,7.34\,kpc) and the outermost ($R >$\,10.34\,kpc) parts of the disc, the phase spirals are barely visible.  

\subsubsection{Slicing by $\Phi$}
\begin{figure}
\centering
\includegraphics[width=3.5in]{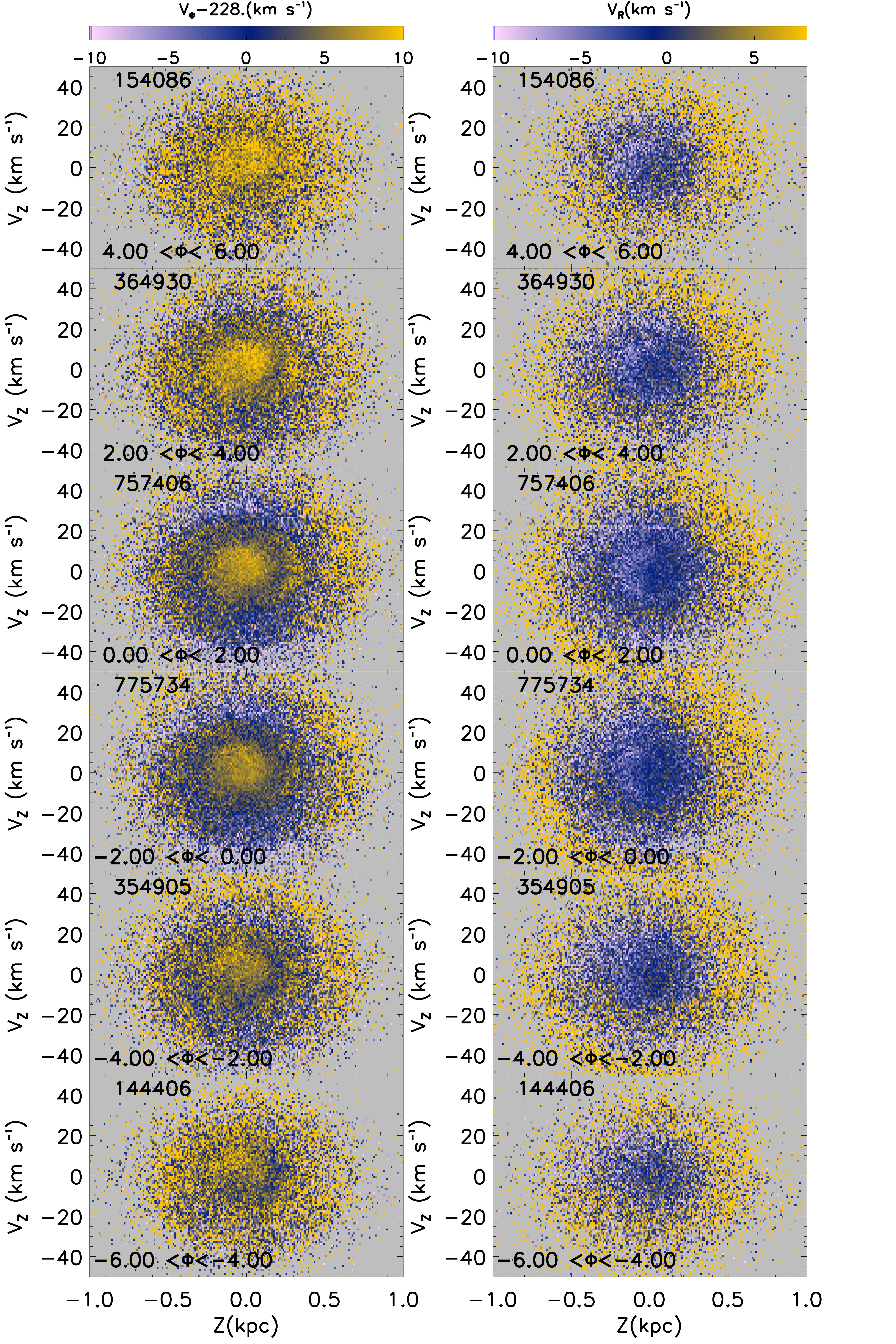}
\caption{ Distributions of median values of $V_{\Phi}-228\,\rm{km s^{-1}}$ (left panels ) and $V_{R}$ (right panels) of the Gaia sample stars  in the $Z$-$V_{Z}$ plane  at different azimuthal angles.  The azimuthal angle range and the number of stars  are labeled in each of the panels. }
\label{low_alpha_phi}
\end{figure}

We now  investigate the distributions of $V_{\Phi}$ and $V_{R}$ of the Gaia sample stars in phase space plane  in different azimuthal  bins. Fig.\,\ref{low_alpha_phi} shows the main results. The azimuthal  range and the number of stars in each bin are labeled in the Figure.  

The phase spirals of $V_{\Phi}$  are apparent in all azimuthal bins. Three spirals are clearly seen, with  shapes quite similar.  The phase spirals become stronger as  $\Phi$ increases, and are the strongest at $-2^{\circ} < \Phi <  2^{\circ}$, and then fade away as $\Phi$ further increases.  Interestingly, the mean values of  $V_{\Phi}$ at $-2^{\circ} < \Phi <  2^{\circ}$ are the smallest. 

The distributions of $V_{R}$  in the phase space plane in the different  $\Phi$ bins are also presented in Fig.\,\ref{low_alpha_r}. Similarly, the phase spirals are found in all the $\Phi$ bins, and are the strongest at $-2^{\circ} < \Phi <  2^{\circ}$. 
Besides, quadrupole patterns are found in all the bins, and are also the strongest at $-2^{\circ} < \Phi <  2^{\circ}$. 

\subsection{The phase spirals at different Galactic  positions as revealed by the Gaia-LAMOST sample}
\begin{figure}
\centering
\includegraphics[width=3.5in]{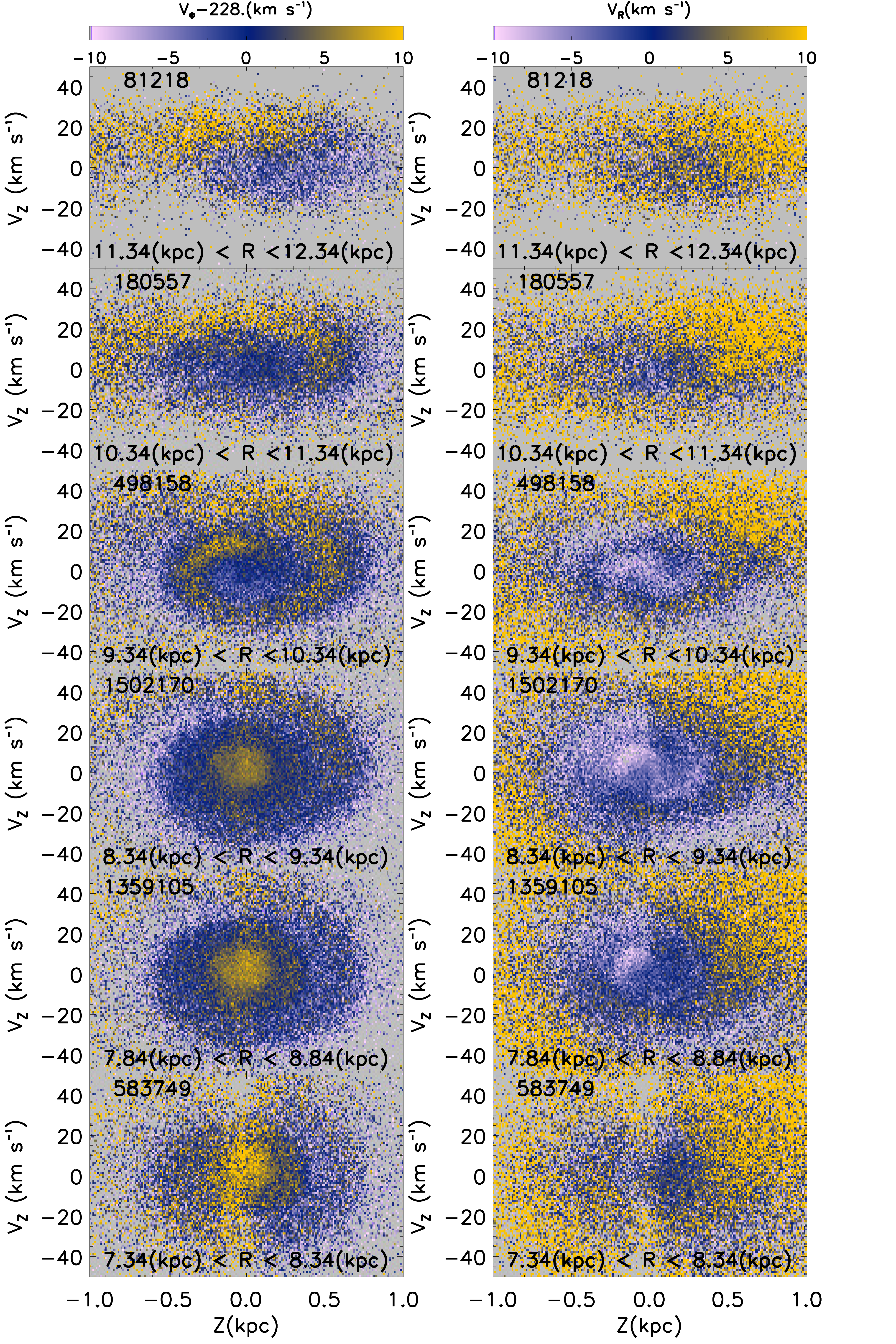}
\caption{ Similar to Fig.\,\ref{low_alpha_r} but for the Gaia-LAMOST sample. The radial range is from 7.34\,kpc to 12.34\,kpc.  }
\label{high_alpha}
\end{figure}

In order to verify  the robustness of our detections of the phase spirals at different Galactic positions, we show in Fig.\,\ref{high_alpha} the $V_{\Phi}$ and $V_{R}$ distributions of the Gaia-LAMOST sample in the $Z$-$V_{Z}$ plane  at different Galactic radii.  What Fig.\,\ref{high_alpha} reveals are broadly similar to those by Fig.\,\ref{low_alpha_r}.

Compared to Fig.\,\ref{low_alpha_r}, the spiral patterns of $V_{\Phi}$ are less well resolved by the Gaia-LAMOST sample stars,  especially for the inner spirals in the inner disc ($7.34 < R < 9.34$\,kpc).  As mentioned in Section\,2,  line-of-sight velocities of the Gaia-LAMOST sample stars  are much less accurate than those of the Gaia sample. This is responsible for the aforementioned difference.  In Fig.\,\ref{high_alpha}, the phase spirals  in outer disc ($R > 9.34$\,kpc) are much stronger than those seen in Fig.\,\ref{low_alpha_r}.  This may be partly due to the fact that  stars  in the Gaia-LAMOST sample have smaller mean values of the absolute azimuthal angle compared to those of the Gaia sample stars at $R > 9.34$\,kpc,  consistent with the results of Section\,3.1.2.

In Fig.\,\ref{high_alpha}, the distributions of $V_{R}$  in the phase space plane  are broadly consistent with what seen  in Fig.\,\ref{low_alpha_r}, but less clearly.  The quadrupole patterns are also similar to those uncovered by the Gaia sample.  In Fig.\,\ref{high_alpha}, the quadrupole patterns  are seen even at $R > 10.34$\,kpc. This is not the case when  using Gaia sample.  The reason is that stars in the  Gaia-LAMOST sample  have smaller mean values of $|\Phi|$ than those of the Gaia sample, considering that the spiral and quadrupole patterns of stars of smaller $|\Phi|$ are stronger  than those of  larger $|\Phi|$. Moreover, there is a clear break of $V_{R}$ from south side ($Z < 0$\,kpc) to  north ($Z\,>\,0$\,kpc), especially in the range of  $7.84 < R < 9.84$\,kpc. We suggest that it might be caused by  the quadrupole patterns of the spirals, the bright limiting magnitudes and very low sampling rates of the LAMOST surveys  in the solar neighbourhood after careful check.

\section{Discussion}
\subsection{The effects of  three dimensional velocity errors  on the phase spirals}
As discussed in Section\,3.1.1, there are 3 phase spirals of  $V_{\Phi}$  in the inner disc ($R < 9.34$\,kpc) but only 1 or 2 phase spirals in the outer disc ($R > 9.34$\,kpc).  Meanwhile, the errors of the three dimensional velocities of stars at $R > 9.34$\,kpc are larger than those of stars at $R < 9.34$\,kpc.  In order to check the relationship between spiral numbers and velocity errors,  we use Monte Carlo method to check if the spirals in the range of [7.84,8.84]\,kpc are still robust. 
We firstly increase the errors of three dimensional velocities of each star at $7.84 < R < 8.84$\,kpc by scaling with a factor, which is  estimated as the ratio of the mean random velocity errors of stars at $9.34 < R < 10.34$\,kpc and those of stars at  $7.84 < R < 8.84$\,kpc. The final velocities of the star are derived by adding the errors, which are randomly generated assuming Gaussian  distributions, with the increased errors as the dispersions. We then examine the resultant  $V_{\Phi}$ and $V_{R}$ distributions in the phase plane. The process is repeated 1000 times, and the resultant 1000 distributions of   $V_{\Phi}$ and $V_{R}$  are quite similar one  to the other.

\begin{figure}
\centering
\includegraphics[width=3.5in]{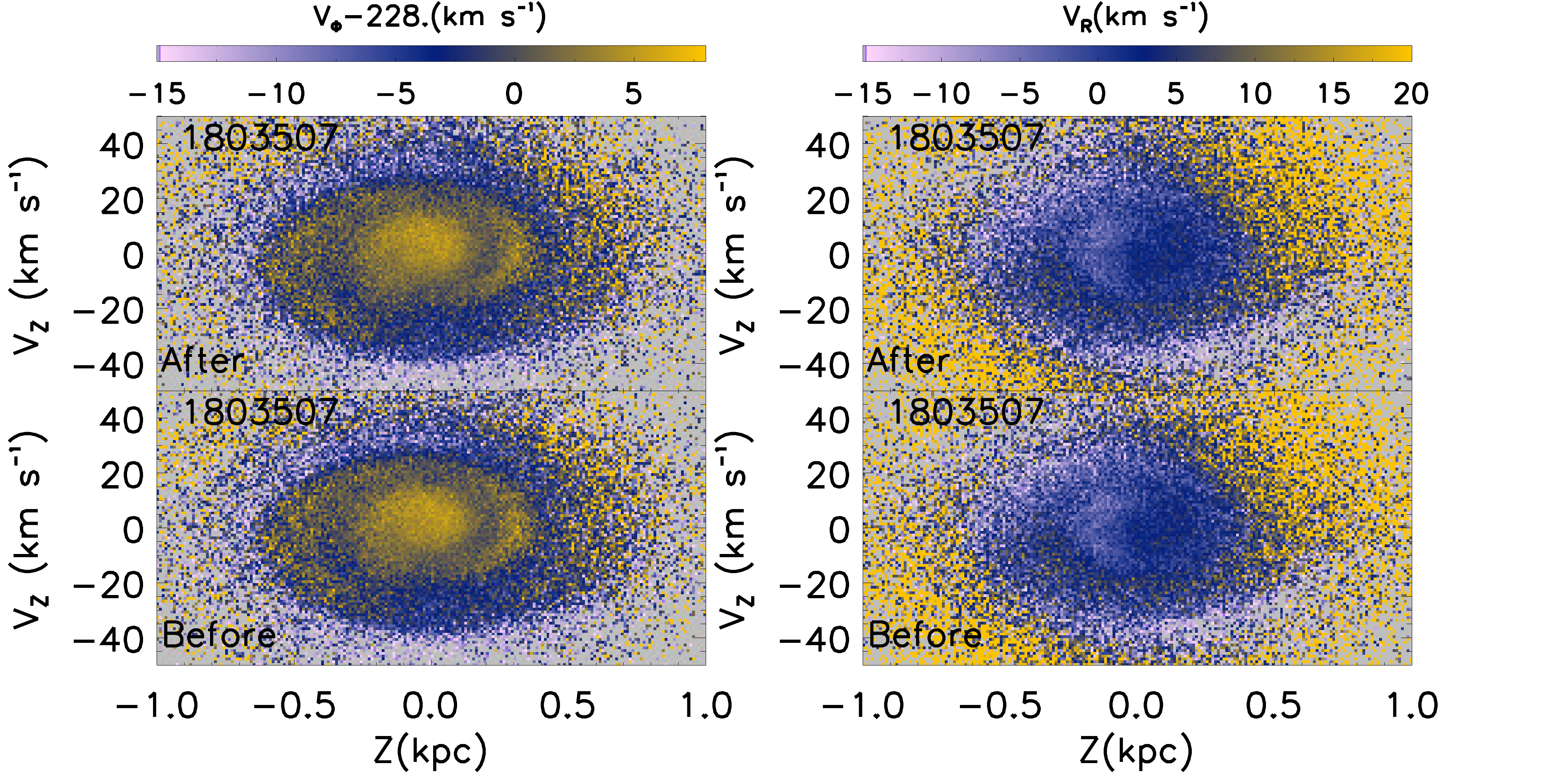}
\caption{Distributions of   $V_{\Phi}-228\,\rm{km s^{-1}}$ (left panels) and $V_{R}$ (right panels)  in the $Z$-$V_{Z}$ plane for stars at  $7.84 < R < 8.84$\,kpc, before (bottom panels) and after (top panels) increasing the velocity errors.}
\label{error}
\end{figure}

Fig.\,\ref{error} shows one of the resultant $V_{\Phi}$ and $V_{R}$ distributions  in the phase space plane  before and after increasing the velocity errors for stars at  $7.84 < R < 8.84$\,kpc.   The phase spirals of $V_{\Phi}$ and $V_{R}$ seen show negligible differences before and after  increasing the velocity errors.  We therefore conclude that the results of more relaxed phase spirals of $V_{\Phi}$ in the outer disc are authentic. 

\subsection{Origins of phase spirals}
Our results as revealed by both the Gaia  and the Gaia-LAMOST samples suggest that the phase space spirals of $V_{\Phi}$ become more relaxed as $R$ increases. The phenomenon is exactly predicted by the  N-body simulations of  \cite{joss}, who suggest that the passage of  the Sagittarius through the Galactic plane, a scenario of an external origin, produces the vertical phase-mixing and phase spirals.
The predicted radial variations of the phase spirals of  $V_{\Phi}$ as driven by buckling of the stellar bar  \citep[internal origin;][]{Khoperskov} are on the other hand  not so consistent with what observed here. 
Our results suggest instead an origin of the  phase spirals that are  developed from  an external perturbation. 

If this scenario of  an external origin is true, the  location of the intruder passing through the Galactic plane and the mass and passage duration of the intruder  become important issues.  
 An intruder will  attract stars to it  changing their three dimensional velocities as the intruder  pass through the Galaxy  \citep[see their Figs.\,1 and 2]{binney2018}. Meanwhile, the phase spirals at different Galactic positions especially its radial and azimuthal variations are no doubt strongly affected by the aforementioned parameters of the intruder.  In other words, one can study the intruding passage location and duration through  the radial and azimuthal variations of the phase spirals of $V_{\Phi}$ and $V_{R}$. 
 
When an intruder pass through the  Galactic plane,  the effect of the intruder on the star in the line of the Galactic center and the intruder is largest, and the phase spiral in this direction is clearest.  In Fig.\,\ref{low_alpha_phi}, we find that the  phase spirals of $V_{\Phi}$ and $V_{R}$ are clearest at $-2^{\circ} < \Phi < 2^{\circ}$, become weaker as $|\Phi|\,>\,2^{\circ}$.  It suggests that the external intruder passes through the Galactic plane in the direction of the Galactic center or anti-center.   


The phase spirals of $V_{\Phi}$ and $V_{R}$ at $4^{\circ} < \Phi < 6^{\circ}$ and  $-6^{\circ} < \Phi < -4^{\circ}$  are significantly weaker, almost invisible, suggesting that the speed of the intruder  is quite fast such that it  mainly affects  stars within an azimuthal angle range of  $8^{\circ}$, or about 2\,per\,cent of the rotation period of a star at a Galactic radius  $R \approx R_{0}$.   The volume that  the intruder can affect also directly linked to its mass, more massive intruder can affect larger volume. 

 As discussed in the previous works, the most probably perturbation source is the Sagittarius dwarf, which has a "trefoil" orbit  over the past 2.3 Gyr. It crossed the disc about 420 Myr ago at $R = 13$\,kpc and transited again about 50 Myr ago  \citep{joss, Law2005,Tepper2018}. If the Sagittarius dwarf is responsible for the phase spirals, the results presented here can help  constrain its orbit 
 and mass, even the shape of the Galactic halo. Detailed modeling is required to reproduce the observed spatial variations of the phase spirals of $V_{\Phi}$ and $V_{R}$.

\section{Summary}
In this Letter, we study the phase space spirals of $V_{\Phi}$ and $V_{R}$ using the Gaia and LAMOST data, and investigate the radial and azimuthal variations of the spiral and quadrupole patterns in the phase space plane.  The main results  are  the following:
\begin{itemize}
\item The distributions of $V_{\Phi}$ in the $Z$-$V_{Z}$ plane show strong spiral patterns at $6.34 < R < 12.34$\,kpc. 
\item The phase spirals of $V_{\Phi}$ become more relaxed as $R$ increases. 
\item The distributions of $V_{R}$ in the $Z$-$V_{Z}$ plane show strong quadrupole patterns in all $R$ and $\Phi$ bins. They also show spiral patterns but  not so tightly wound as those of $V_{\Phi}$. In the innermost ($R <$ 7.34 kpc) and outermost disc ($R >$ 10.34 kpc), the spiral patterns are barely visible. 
\item The spiral patterns of $V_{\Phi}$  and the quadrupole and spiral patterns of $V_{R}$ in the $Z$-$V_{Z}$ plane are the strongest at $-2^{\circ} < \Phi < 2^{\circ}$, but almost invisible at $4^{\circ} < \Phi < 6^{\circ}$ and $-6^{\circ} < \Phi < -4^{\circ}$. 
\end{itemize}

The radial variations of the phase spirals of $V_{\Phi}$  is consistent with the predictions of perturbation by an  external intruder as suggested by   \cite{joss}, but inconsistent with an  internal origin as  suggested by  \cite{Khoperskov}. The azimuthal variations of the phase spirals of $V_{\Phi}$ and $V_{R}$   suggest that the intruder passes through the Galactic plane in the direction of either the  Galactic center or  the anti-center. The azimuthal variations of the phase spirals of $V_{\Phi}$ and $V_{R}$ also tell us something about  the mass  and the passage duration of the intruder. A detailed model is required to reproduce the results presented here. 

\section{ACKNOWLEDGEMENTS}
 This work is supported by   the National
Natural Science Foundation of China 11833006, U1531244 and 11473001.  Guoshoujing Telescope (the Large Sky Area Multi-Object Fiber Spectroscopic
Telescope LAMOST) is a National Major Scientific Project built by the Chinese Academy of Sciences.
Funding for the project has been provided by the National Development and Reform Commission.
LAMOST is operated and managed by the National Astronomical Observatories, Chinese Academy of
Sciences. The LAMOST FELLOWSHIP is supported by Special Funding for Advanced Users, budgeted and administrated by Center for Astronomical Mega-Science, Chinese Academy of Sciences (CAMS).

\bibliographystyle{apj}

\bibliography{phase_spiral}
\end{document}